\documentclass[a4paper,
               biblatex,     % biblatex is used
               ]{jacow}
\usepackage{pdfpages,multirow,ragged2e} %
\usepackage{graphics}
\makeatletter%
	\ifboolexpr{bool{xetex}}
	 {\renewcommand{\Gin@extensions}{.pdf,%
	                    .png,.jpg,.bmp,.pict,.tif,.psd,.mac,.sga,.tga,.gif,%
	                    .eps,.ps,%
	                    }}{}
\makeatother

\ifboolexpr{bool{xetex} or bool{luatex}} % test for XeTeX/LuaTeX
 {}                                      % input encoding is utf8 by default
 {\usepackage[utf8]{inputenc}}           % switch to utf8

\usepackage[USenglish]{babel}

\ifboolexpr{bool{jacowbiblatex}}%
 {%
  \addbibresource{MOPS74.bib}
  \addbibresource{biblatex-examples.bib}
 }{}
\begin{document}

\title{Accelerator system parameter estimation using variational autoencoded latent regression \thanks{Work supported by the LANL LDRD Program Directed Research (DR) project 20220074DR}}

\author{M. Rautela \thanks{mrautela@lanl.gov}\textsuperscript{1}, A. Williams\textsuperscript{1,2}, A. Scheinker\textsuperscript{1} \\ 
        \textsuperscript{1}Los Alamos National Laboratory, NM, US \\
        \textsuperscript{2}University of California, San Diago, California, US}
	
\maketitle

\begin{abstract}
Particle accelerators are time-varying systems whose components are perturbed by external disturbances. Tuning accelerators can be a time-consuming process involving manual adjustment of multiple components, such as RF cavities, to minimize beam loss due to time-varying drifts. The high dimensionality of the system ($\sim$100 amplitude and phase RF settings in the LANSCE accelerator) makes it difficult to achieve optimal operation. The time-varying drifts and the dimensionality make system parameter estimation a challenging optimization problem. In this work, we propose a \textbf{V}ariational \textbf{A}utoencoded \textbf{L}at\textbf{e}nt \textbf{R}egression (VALeR) model for robust estimation of system parameters using 2D unique projections of a charged particle beam's 6D phase space. In VALeR, VAE projects the phase space projections into a lower-dimensional latent space, and a dense neural network maps the latent space onto the space of system parameters. The trained network can predict system parameters for unseen phase space projections. Furthermore, VALeR can generate new projections by randomly sampling the latent space of VAE and also estimate the corresponding system parameters.
\end{abstract}

\section{INTRODUCTION}
%Particle accelerators are complex machines which are used to focus, guide and accelerate charged particles to high energy for various scientific applications. 
Charged particle dynamics are governed by hundreds to thousands of accelerator components and during operation these system parameters are adjusted manually to achieve minimal beam loss. The process is time-consuming and leads to suboptimal performance due to the time-varying nature of the system. % parameters introduces additional uncertainties. %In order to precisely estimate, tune and optimize beam parameters, understanding beam dynamics plays an important role. 
Detailed non-destructive beam measurements are limited at most accelerators, making virtual beam diagnostics attractive. The first approach to virtual beam diagnostics coupled an online physics model \cite{tenenbaum2005lucretia} with non-invasive beam measurements through adaptive feedback \cite{scheinker2021extremum} to track the time-varying longitudinal phase space of the electron beam in FACET at SLAC \cite{scheinker2015adaptive}. These initial virtual beam diagnostics results inspired various ML-based approaches to the same problem \cite{emma2018machine,zhu2021high}. 

Physics-based simulations can be computationally expensive \cite{young2003particle,pang2014gpu}, which makes model-based approaches to inverse problems, parameter estimation, tuning, control, and optimization challenging \cite{newton1970inverse}. Machine learning (ML)--based approaches on modern GPUs are orders of magnitude faster. 

Recently the use of advanced adaptive and ML-based methods for tuning and optimization in particle accelerators has grown in popularity. In \cite{scheinker2018demonstration}, the first approach to adaptive ML was demonstrated by combining a deep neural network with model-independent feedback for automatic control of the longitudinal phase space of intense electron beams in the LCLS FEL. In \cite{li2018genetic,wan2019improvement}, the K-means clustering algorithm was introduced to improve the multi-objective genetic algorithm (MOGA) for optimization of the NSLS-II storage ring’s dynamic aperture. In \cite{scheinker2020online} real-time multiobjective optimization was demonstrated on the AWAKE electron beamline for simultaneous emittance and beam orbit control. In \cite{huang2019multi}, a multi-objective multi-generation Gaussian process regression (GPR) model is proposed for design optimization. In \cite{edelen2020machine}, neural network surrogates are used for MOGA simulation studies of accelerators.

In \cite{wan2020neural}, neural network-based MOGA is proposed for Touschek lifetime and dynamic aperture optimization of the high energy photon source. In \cite{kirschner2022tuning}, safe Bayesian optimization (BO) is introduced for online tuning of accelerators. BO creates a probabilistic model of the objective function and selects the next evaluation point to maximize the acquisition function, while balancing exploration and exploitation \cite{rautela2023bayesian}. Reinforcement learning is also utilized for the optimization of accelerators \cite{kaiser2022learning,meier2022optimizing}. In \cite{hwang2022prior}, neural networks trained on historic data are utilized as priors for GP-based Bayesian optimization and demonstrated for tuning the FRIB front-end. ML is also useful for building inverse models which can greatly benefit the optimization process \cite{rautela2022inverse}. Inverse models have been developed to map downstream measurements to initial beam conditions \cite{scheinker2021adaptive_SciRep}. 

In this paper, we introduce \textbf{V}ariational \textbf{A}utoencoded \textbf{L}at\textbf{e}nt \textbf{R}egression (VALeR) model for accelerator system parameter estimation (RF cavity settings) given six-dimensional phase space of charged particle beams. VALeR consists of a variational autoencoder-based generator and a dense neural network (DNN) based regressor. VAE projects 15 unique projections of 6D phase space into a lower dimensional continuous latent representation and the DNN maps the latent space to the accelerator settings. This novel coupled generator-regressor model predicts RF settings for unseen phase space projections. The model also generates new realistic projections as well as their RF settings by sampling the latent space followed by decoding and regression, respectively. 

\section{METHODS}
%\subsection{Multi-particle tracking simulations}
%This investigation centers on the LANSCE linear accelerator at Los Alamos National Laboratory \cite{wangler2008rf}. LANSCE accelerates high-intensity $H^+$ and $H^-$ beams to 800 MeV for various experiments including ultracold neutron science, proton radiography, and neutron scattering \cite{scheinker2021extremum}. Following bunching in the LEBT, the LANSCE beam enters the 201.25-MHz Drift Tube linac (DTL) for acceleration to 100 MeV. The DTL is segmented into four modules, each powered by a distinct RF amplifier and housing numerous focusing quadrupole magnets. The beam is then accelerated up to 800 MeV in a 805 MHz coupled-cavity linac (CCL). The CCL contains an additional 44 modules, each featuring an independently driven accelerating structure and quadrupole doublets along its 726-meter length.

This work centers on the LANSCE linear accelerator at Los Alamos National Laboratory. Details about LANSCE tuning challenges are given in \cite{scheinker2021extremum}. During accelerator operation, adjustments to magnet and RF settings are usually tuned manually to minimize beam loss. This method is time-consuming, often failing to attain optimal performance. To achieve optimal functionality, understanding beam dynamics is crucial. Various simulation tools have been devised for this purpose \cite{tenenbaum2005lucretia,young2003particle,pang2014gpu}. High Performance Simulator (HPSim) is an advanced multiple-particle beam dynamics simulator taking into account external accelerating and focusing forces as well as space charge forces \cite{pang2014gpu}. We generate synthetic data by randomly sampling RF set points (amplitude and phase) of the first four modules from a uniform distribution keeping other beam and accelerator parameters fixed. 

\begin{figure}
    \centering
    \begin{minipage}[b]{1.0\linewidth}
        \centering
        \includegraphics[width=1.0\textwidth]{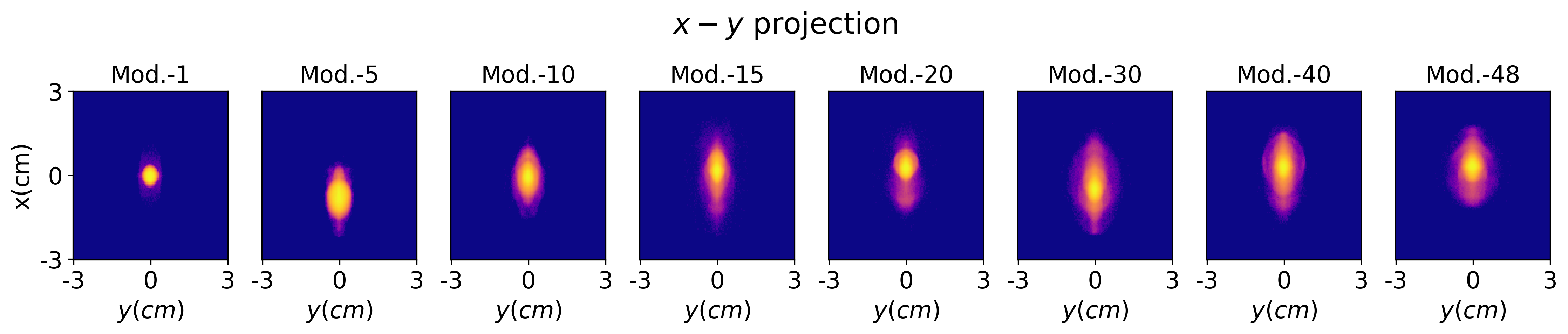}
    \end{minipage}
    \begin{minipage}[b]{1.0\linewidth}
        \centering
        \includegraphics[width=1.0\textwidth]{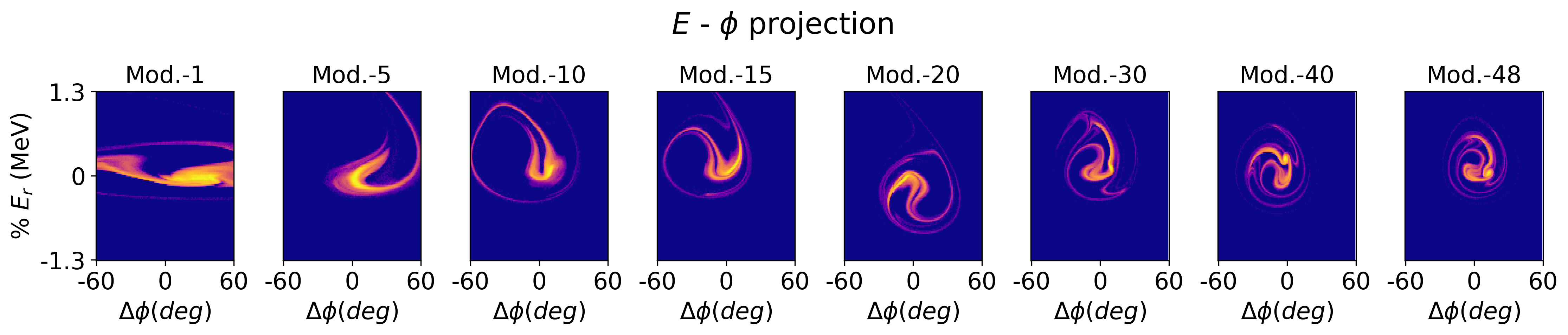}
    \end{minipage}
    \begin{minipage}[b]{1.0\linewidth}
        \centering
        \includegraphics[width=1.0\textwidth]{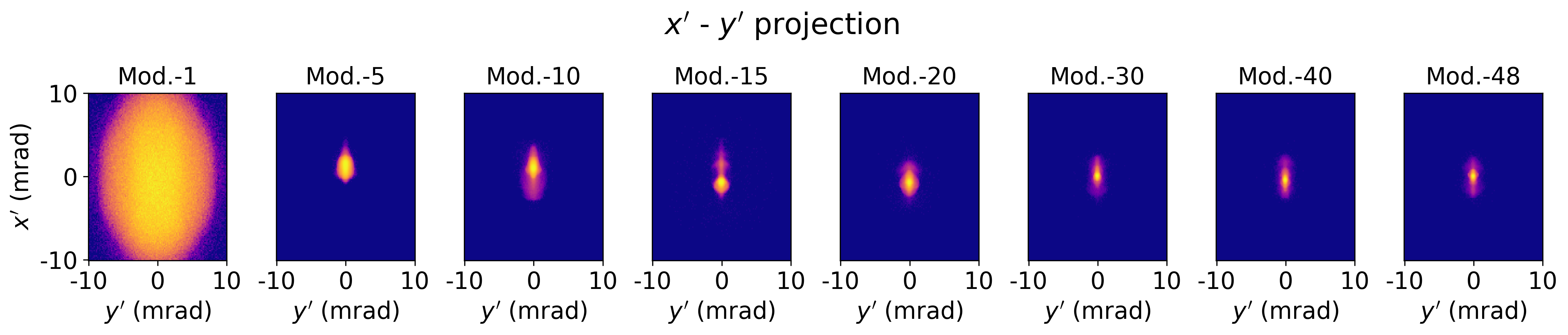}
    \end{minipage}
    \caption{Three out of fifteen projections ($x-y$, $E-\phi$, $x'-y'$) of the 6D phase space of charged particle beams at different modules. The plots are shown on a logarithmic scale for better visualization.}
    \label{fig:dataset}
\end{figure}

From the HPSim output, we generate 2D histograms which are the 15 unique projections of the beam's 6D $(x,y,z,p_x,p_y,p_z)$ phase space at each of the 48 modules. For our application $(z,p_z)$ is converted to $(\phi,E)$ where $\phi$ is the phase of a particle in a bunch relative to the design phase. In Fig.~\ref{fig:dataset}, 3 of the 15 phase space projections ($x,y$, $E,\phi$, and $x',y'$) are plotted at various accelerating modules.

\begin{figure}
    \centering
    \includegraphics[width=0.5\textwidth]{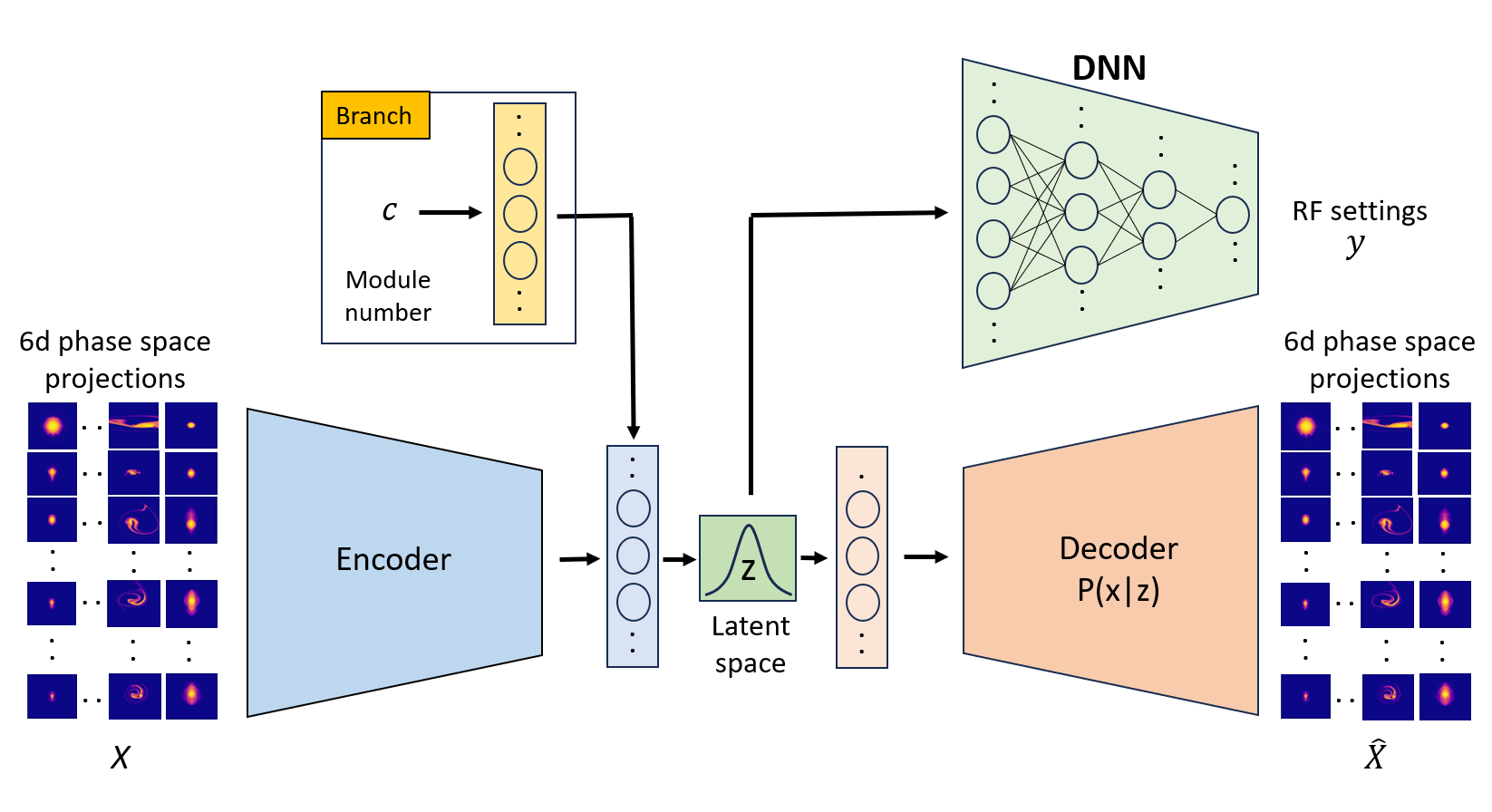}
    \caption{Variational autoencoded latent regression (VALeR) model. A conditional VAE transform phase space ($X$) and module numbers ($c$) into a latent space distribution ($z$). A DNN maps the latent space to the RF settings ($y$).}
    \label{fig:vaelr}
\end{figure}

%\subsection{VALeR model}
Autoencoders (AE) are able to learn low-dimensional latent representations of complex data and then generate new high-dimensional data from the latent embedding \cite{rautela2022delamination}. Recently, adaptive latent space tuning methods have been developed using conditional generative AEs with compact latent space representations that can be smoothly traversed to generate the 15 projections of a beam's 6D phase space for beam diagnostics \cite{scheinker2021adaptive_SciRep, scheinker2023adaptive_PRE}. Variational autoencoders (VAE) are AEs that map to a probabilistic latent space \cite{rautela2022towards}. VAEs also enable the generation of new realistic samples by traversing the latent space \cite{rautela2024conditional}. In this work, we propose a VALeR model that utilizes a conditional VAE to transform 15 unique projections of 6D phase space into a lower dimensional latent space and a DNN maps the latent space to the accelerator settings. The architecture of the VALeR model is shown in Fig.~\ref{fig:vaelr}.

The relationship between phase space ($X$) and accelerator settings ($y$) can be captured by a directed acyclic graph (DAG) or a Bayesian network, i.e., $X \rightarrow y$ \cite{heckerman2008tutorial}. The joint probability distribution can be written as $P(X,y) = P(y|X)$ and in deterministic settings as $f:X\rightarrow y$. A neural network $NN(W,b)$ with an encoder type of architecture can be trained to learn this functional relationship. However, in our research, we introduce a latent variable $z$, so that the DAG becomes $X \rightarrow z \rightarrow y$. Therefore, $P(y|X) = P(z|X) P(y|z)$, where $P(z|X)$ and $P(y|z)$ can be learned with a VAE and a DNN, respectively. This unique way of introducing two neural networks to map $X$ to $y$ instead of one brings multiple advantages. The latent space of a VAE can be sampled to generate new realistic X and y, which is not possible with a simple encoder type of neural network. While trained on limited data (X and y), this surrogate simulator can generate realistic unlimited data in a short period of time, which is practically impossible with the physics-based simulator. The generation ability for uncertainty analysis, control, and design of accelerators is the subject of future work.

\section{RESULTS}
To generate data, the RF field set points (amplitude and phase) of the DTL sections (first 4 modules) were randomly sampled (3300 simulations) while keeping the other 88 RF parameters fixed. HPSim would then simulate the dynamics of a beam through the entire accelerator, from which the 15 unique phase space projections (each with a 128 $\times$ 128 image) at each of the 48 RF modules were generated. 3200 simulation data sets were used for training and 100 for testing. A single input to the VAE is a set of 15 128 $\times$ 128-pixel images. The conditional input $c$ to the encoder is the module number, a scalar between 1-48, normalized to the range $[0,1]$.  

%The initial 3200 data objects were divided into training and validation sets with a 85:15 ratio. The CVAE utilizes a symmetrical encoder-decoder network with five convolutional layers with 32, 64, 128, 256, 512 filters of size 3$\times$3 each with strides of 2, followed by a dense layer with 256 neurons. The branch network with two layers of 32 neurons is concatenated with the primary VAE architecture and transformed into a 8 dimensional latent space. The activation function is set to LeakyReLU, followed by batch normalization for every layer. The learning rate of 0.001 and a batch size of 32 is found to perform well. The CVAE network is trained with the Adam optimizer for 1500 epochs with ELBO loss \cite{rautela2024conditional}. The latent space of CVAE is reshaped into 348 features (modules $\times$ latent dimension) and is used as input to the DNN and RF settings as output. The DNN has 5 layers with 1024, 1024, 1024, 512, 128 neurons and is trained with mean squared loss function for 5000 epochs. Other hyper-parameters remain similar to CVAE.

\subsection{Prediction Ability of VALeR}
\begin{figure}
    \centering
    \begin{minipage}[b]{0.49\linewidth}
        \centering
        \includegraphics[width=1.0\textwidth]{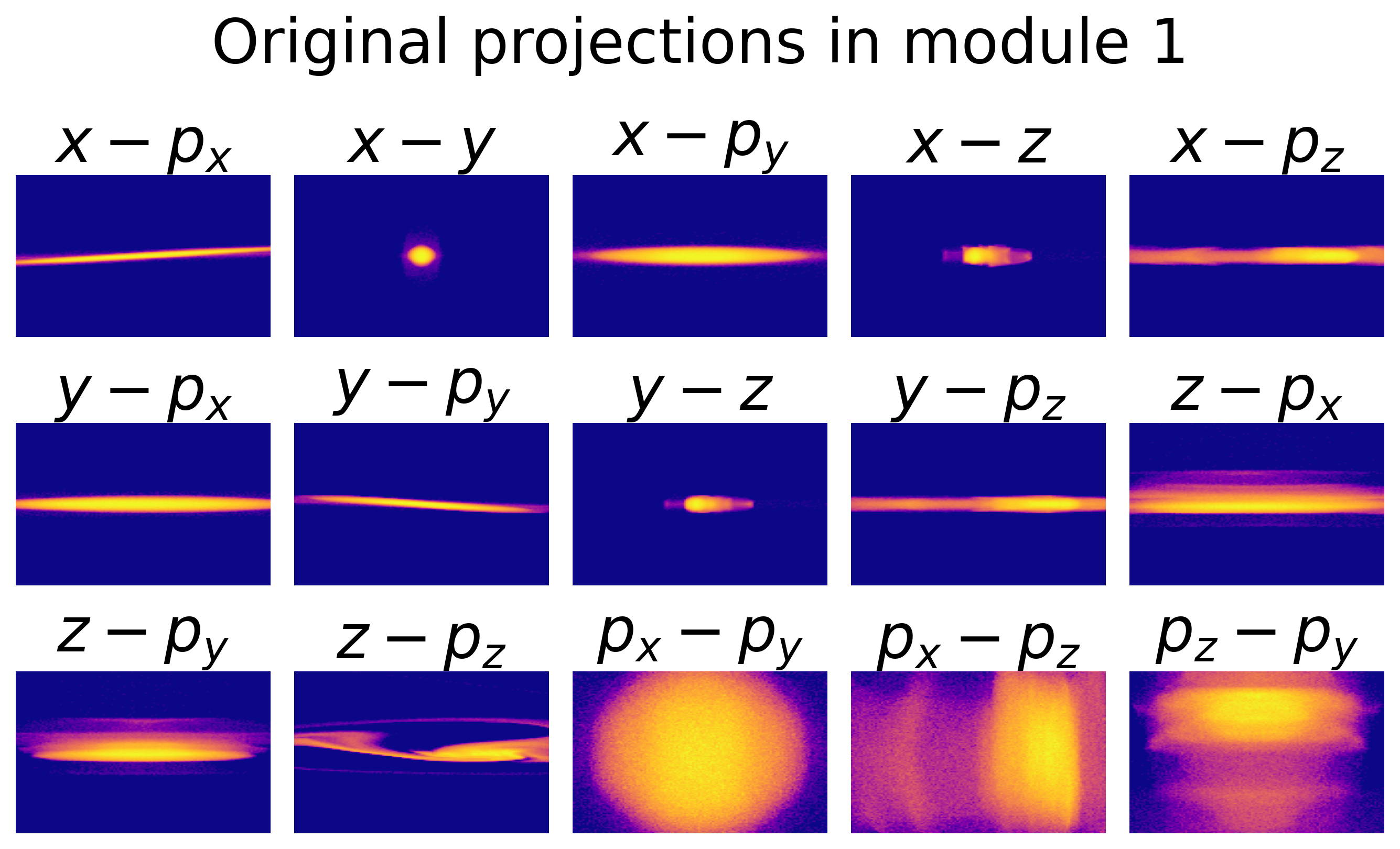}
    \end{minipage}
    \begin{minipage}[b]{0.49\linewidth}
        \centering
        \includegraphics[width=1.0\textwidth]{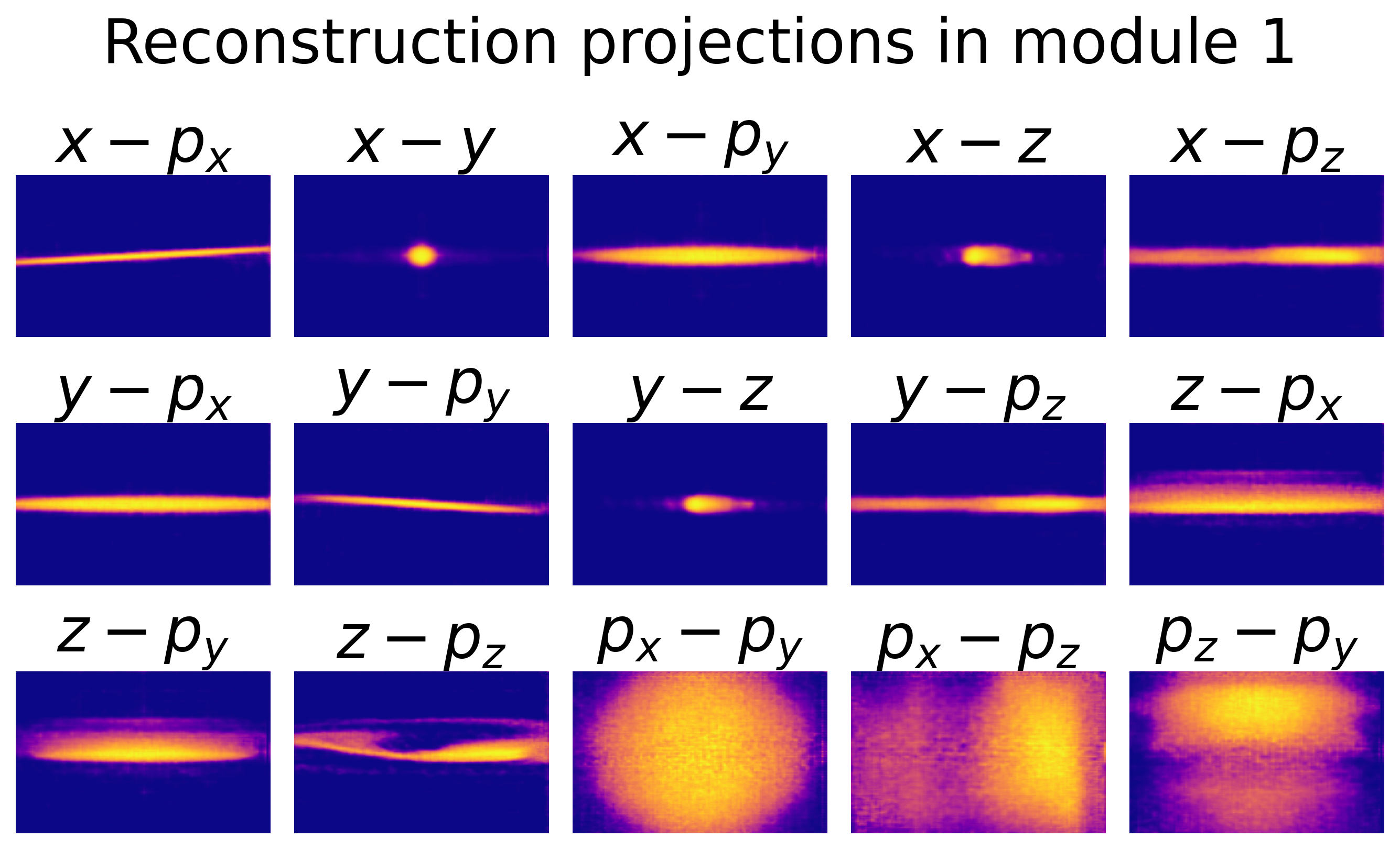}
    \end{minipage}
    \begin{minipage}[b]{0.49\linewidth}
        \centering
        \includegraphics[width=1.0\textwidth]{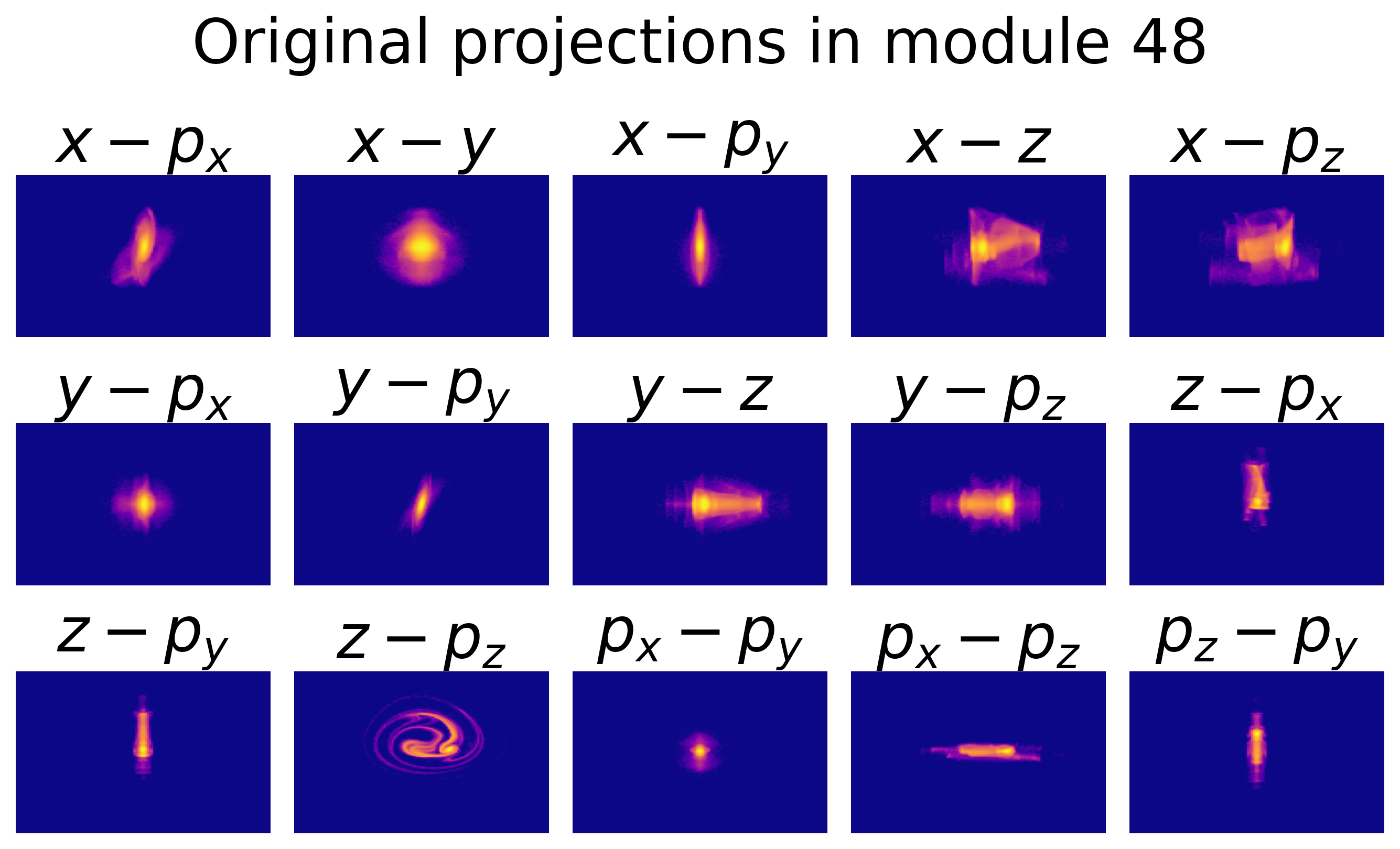}
    \end{minipage}
    \begin{minipage}[b]{0.49\linewidth}
        \centering
        \includegraphics[width=1.0\textwidth]{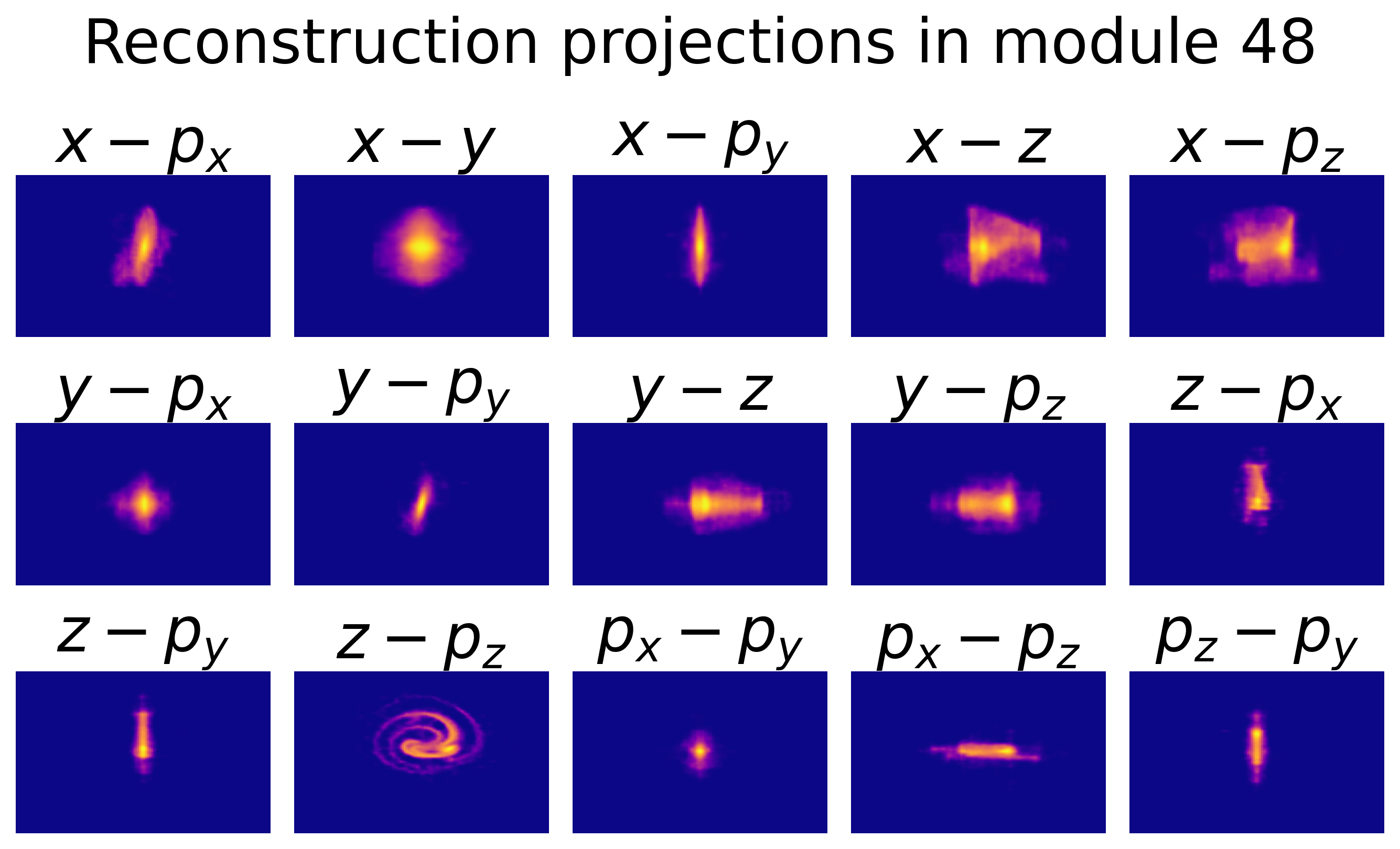}
    \end{minipage}
    \caption{Prediction results: Original vs Reconstructed projections from the test dataset.}
    \label{fig:reconability}
\end{figure}

\begin{figure}
    \centering
    \includegraphics[width=0.49\textwidth]{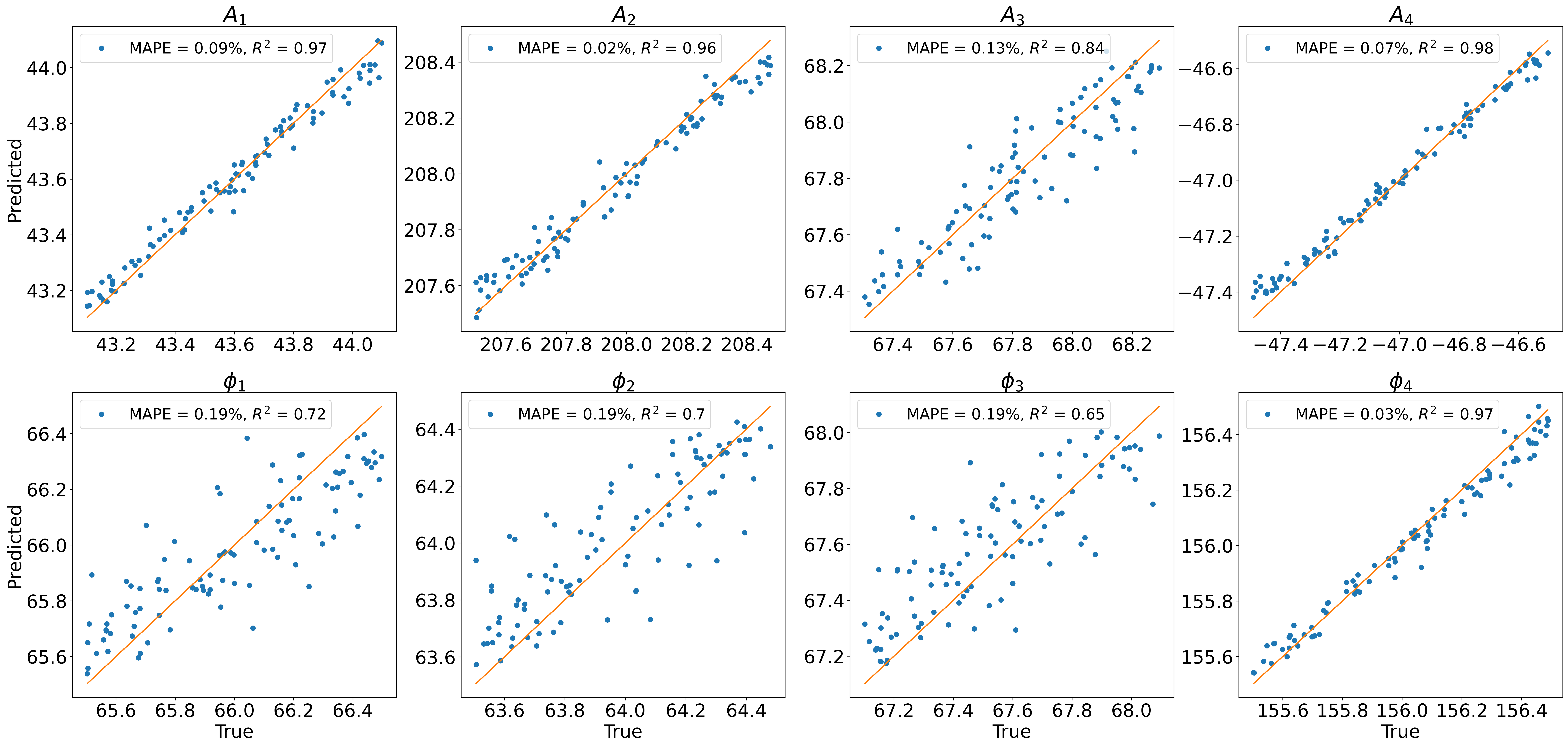}
    \caption{Prediction results: True vs Predicted RF settings.}
    \label{fig:yorgvsyhat}
\end{figure}

The test set is used to study the ability of VALeR to generalize. Figs.~\ref{fig:reconability} and \ref{fig:yorgvsyhat} show the prediction results. In Figs.~\ref{fig:reconability}, the original projections from the test set are plotted against the reconstructions obtained from the CVAE. An average MSE of 5e-7 and structural similarity index (SSIM) of 0.989 is recorded for the images. In Fig.~\ref{fig:yorgvsyhat}, all eight true and predicted RF settings are plotted with their respective $R^2$ and mean absolute percentage error (MAPE). We have recorded MAPE of 0.09$\%$, 0.02$\%$, 0.13$\%$, 0.07$\%$, 0.19$\%$, 0.19$\%$, 0.19$\%$, and 0.03$\%$ which confirms the accuracy of predicted RF settings. The coefficient of determination ($R^2$ value) is also calculated to check the goodness of fit between true and predicted values. We have noted 0.97, 0.97, 0.84, 0.98, 0.72, 0.7, 0.65, and 0.97. The $R^2$ value of phase settings of the first 3 modules i.e., $\phi_1$, $\phi_2$, and $\phi_3$ are not as exceptional as amplitude settings. The reason for this is due to the lower sensitivity of phase settings (as compared to amplitude settings) on the phase space of the beam. Another factor can be the limited dataset used for training the network. In future investigations, we are planning to train the network with more data.

\subsection{Generation Ability of VALeR}
We have performed generation based on random perturbation in a vicinity around the latent point of a test sample. This is different from the Monte-Carlo sampling-based generation where a random point is unconditionally sampled within the bounds of the latent space. With a limited dataset used to train the regressor, the network provides limited mapping capability on unconditionally sampled latent points.

At first, a test sample is projected on the latent space, which transforms phase space projections in all the modules into 48 different 8-dimensional latent points. We introduce an Euclidean vicinity around these points, which is a circle of maximum $\alpha$ radius. The variable $\alpha$ is defined as a percentage of the latent space bounds of that module. We have set $\alpha$ at 2.5$\%$. A random variable $\epsilon \sim U(0,1)$ is defined for random sampling from a uniform distribution. A new latent point is sampled using $z_{new} = z + \alpha \odot \epsilon$, where $\alpha \odot \epsilon$ is the perturbation around the latent point corresponding to the test sample. The sampled point is fed to the decode as well as the DNN regressor to generate new phase space projections as well as their corresponding RF settings. 

\begin{figure}
    \centering
    \begin{minipage}[b]{1.0\linewidth}
        \centering
        \includegraphics[width=1.0\textwidth]{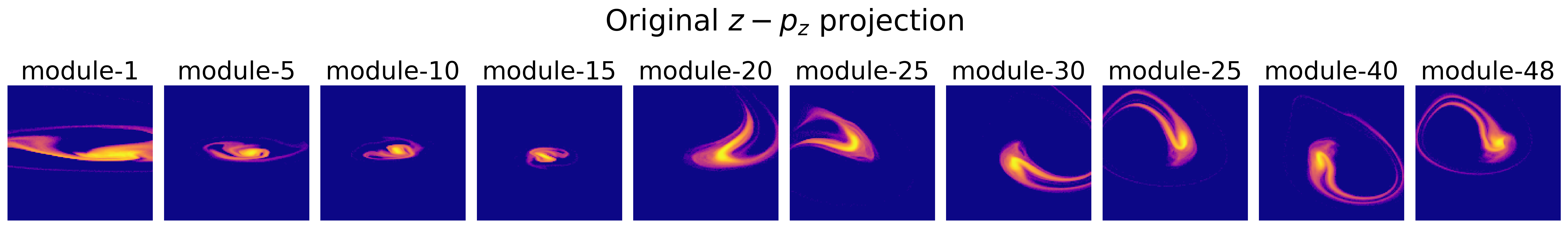}
    \end{minipage}
    \begin{minipage}[b]{1.0\linewidth}
        \centering
        \includegraphics[width=1.0\textwidth]{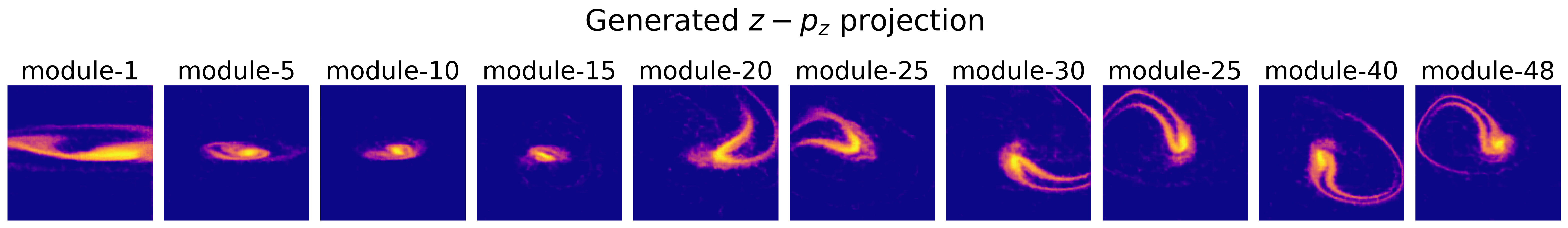}
    \end{minipage}
    \caption{Generation results: original and generated $z-p_z$ projection using CVAE}
    \label{fig:genability}
\end{figure}

We have recorded an average MSE of 6.23e-6 and SSIM of 0.965 for generated projections. One of the generated $z-p_z$ projections for different modules is shown in Fig.~\ref{fig:genability} against the original test sample. The generated RF accelerator settings have mean absolute percentage differences of 0.098$\%$, 0.005$\%$, 0.03$\%$, 0.091$\%$, respectively whereas the generated phase settings have mean absolute differences of 0.295$^{\circ}$, 0.42$^{\circ}$, 0.037$^{\circ}$, 0.016$^{\circ}$, respectively.

The RF settings are within the bounds of the settings used to produce phase space projections from HPSim. However, this is not sufficient to prove the generative ability of VALeR. For further validation, generated RF settings can be input into HPSim, and the simulated phase space projections compared with the VALeR-generated projections as a part of future research.

\section{CONCLUSION}
A novel coupled generator-regressor model, called VALeR is proposed to estimate the RF parameter settings given the phase space of charged particles along the accelerator. A CVAE projects the phase space into the latent space followed by a DNN to map it to the RF settings. In the prediction and generation task, we recorded low MSE and high SSIM on the images as well as low MAPE on the predicted RF settings. We have also seen that the generated RF settings are well within the bounds of the training set. The model can not only be used for tuning and optimization studies but also finds application for the design and control of the accelerators.

\section{ACKNOWLEDGEMENTS}
Work supported by Los Alamos National Laboratory LDRD Program DR project 20220074DR.

\ifboolexpr{bool{jacowbiblatex}}%
	{\printbibliography}%
	{%
	% "biblatex" is not used, go the "manual" way
	
	%\begin{thebibliography}{99}   % Use for  10-99  references
	
}

\end{document}